\documentclass{article}
\usepackage{psfig}
\begin{document}
\title{The Fermi surfaces of Metallic Alloys and the Oscillatory Magnetic
Coupling between Magnetic Layers separated by such Alloy Spacers}

\author{Gy\"orffy, B. L. 
and Lathiotakis N. N. \\
{\it  H. H. Wills Physics Laboratory, University of Bristol,} \\
{\it Royal Fort, Tyndall Avenue, Bristol BS8~1TL, U.K.} 
}
\date{\today}
\maketitle
\begin{abstract}
We review the theory of oscillatory magnetic coupling in Metallic Multilayers
across alloy spacers. We illustrate the relationship between the frequencies of
the oscillations and the extremal caliper vectors of the Fermi surface of the 
spacer by explicit calculations for Cu$_{(1-x)}$Ni$_x$, Cr$_{(1-x)}$V$_x$ and
Cr$_{(1-x)}$Mo$_x$ alloys. We argue the measurement of the frequencies of such
oscillations can be an extremely useful and cheap probe of the Fermi surface of
random alloys. 
\end{abstract}

\section{Introduction}
Many random alloys such as Cu$_{(1-x)}$Ni$_x$, Cu$_{(1-x)}$Au$_x$ are metals and
therefore have Fermi surfaces in a well defined sense\cite{gyorffy1}. Moreover
these Fermi surfaces determine many of the properties of these scientifically 
interesting and technologically important class of materials. Thus, it would be
useful to know what these Fermi surfaces are like and how they evolve with
changing concentration. Unfortunately the classic probes of the Fermi surface
such as the measurement of the de Haas van Alphen (dHvA) 
oscillations\cite{dhva,coleridge} work only if the quasi-particles can complete a cycle
along the Landau orbit between two scattering events associated by deviations of
the crystal potential from periodicity. As it happens this physical requirement
of long quasi-particle life times translates into very small, $\sim$ppm, 
concentration of impurities and hence no dHvA signal is expected for the
concentrated alloys of interest. This leaves, until recently, two dimensional
Angular Correlation of (Positron) Annihilation Radiation (2d ACAR) and Compton
Scattering (SC) studies, which do not require long quasi-particle life times, as
the only source of reasonably direct quantitative information about the Fermi
Surfaces of Random Alloys. Our aim here is to argue that measurements of the
oscillatory coupling between magnetic layers across random alloy spacers can
also provide such information. 

\begin{figure}
\centerline{\psfig{figure=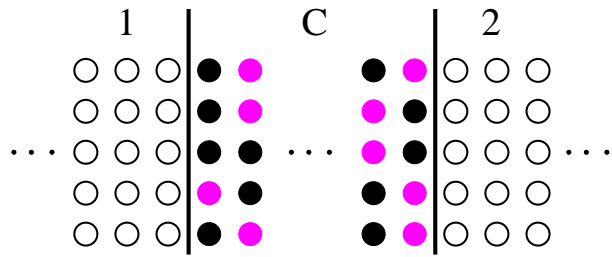,width=8 cm}}
\caption{Schematic view of a sandwich structure of 1 and 2 magnetic layers
separated by a non magnetic spacer layer C which could be 
disordered in general. 
 \label{fig:struct}}
\end{figure}

In short, what one measures is the exchange coupling $J_{12}(L)$ between
magnetic layers 1 and 2 separated by a non magnetic spacer layer of thickness
$L$ made of a random metallic alloy. A schematic picture of the experiment
configuration is shown in Fig.~\ref{fig:struct} and the exchange interaction
$J_{12}$ is defined in terms of the magnetic interaction energy 
$\delta E_{12}(L)$:
\begin{equation}
\delta E_{12} (L) = J_{12}(L) \; {\bf M}_1  {\bf M}_2 
\label{eq:1}
\end{equation}
where ${\bf M}_1$, ${\bf M}_2$ are the average magnetization of magnetic layers
1 and 2 respectively. As was discoverer by Parkin {\it et al}\cite{parkin0}, and
has been observed for a vast variety of systems\cite{review}, $J_{12}(L)$
oscillates between being Ferromagnetic, $J_{12}<0$ and Antiferromagnetic, 
$J_{12} > 0$ as a function of the separation $L$. In fact most experiments seem
to be consistent with the formula\cite{bruno,edwards}
\begin{equation}
J_{12} (L) = - \frac{1}{L^2} \sum_\nu A_\nu \cos (Q_\nu L + \phi_\nu ) \;
e^{-L/\Lambda_\nu} 
\label{eq:2}
\end{equation}
where each contribution $\nu=1,2,\dots$ is characterized by the period
$P_\nu=2\pi /Q_\nu$, amplitude $A_\nu$, phase $\phi_\nu$ and coherence length 
$\Lambda_\nu$ which is infinite for pure metal spacer but is finite if the
spacer is a random metallic alloy. Our discussion will focus on the astonishing
fact that $Q_\nu$ is a quantitative measure of a geometrical feature of the 
Fermi surface of the infinite (bulk) spacer metal. In fact, most theories
predict the form in eq.~(\ref{eq:1}) asymptotically for large $L$ and $Q_\nu$ 
turns out to be an extremal caliper vector, connecting two opposite points on the
Fermi surface, in the direction perpendicular to the plane of the magnetic
layers, that is to say in the growth direction of the multilayer system. This 
result is well confirmed for a large number of pure metal spacers. In the next
two sections we review the agreement between the predicted and measured
evolution of $Q_\nu$ with concentration $x$ for random metallic alloys. In
section~\ref{SEC:4} we will discuss the exponential damping factor and in 
section~\ref{SEC:5} we present an asymptotic theory for the amplitudes and the 
phases $Q_\nu$. 

The interest in the above interlayer magnetic coupling has arisen in the wake of
its discovery, because of its connection with the technologically very important
Giant Magneto-resistance (GMR) phenomenon. From this point of view the problem
is largely solved. The physical mechanism understood to be the planar defect
analogue of the RKKY interaction between point like magnetic defects in metals,
enhanced by confinement\cite{bruno}. In this contribution we wish to 
emphasize an other
aspect of the problem. Namely, we shall explore the possibility of using the
oscillatory coupling phenomenon as a new probe of the Fermi surface in
transition metal alloys. Clearly, compared with 2d-ACAR the measurement of these
oscillations are simple and cheap and hence the prospects of such project are
bright. However, before the full power of the method can be assessed both the
experimental technique and the theoretical framework used to interpret the data
will require further scrutiny.
In what follows we will present a few initial steps in the direction of the
latter. 

\begin{figure}[h]
\centerline{ \begin{tabular}{cc}
\psfig{figure=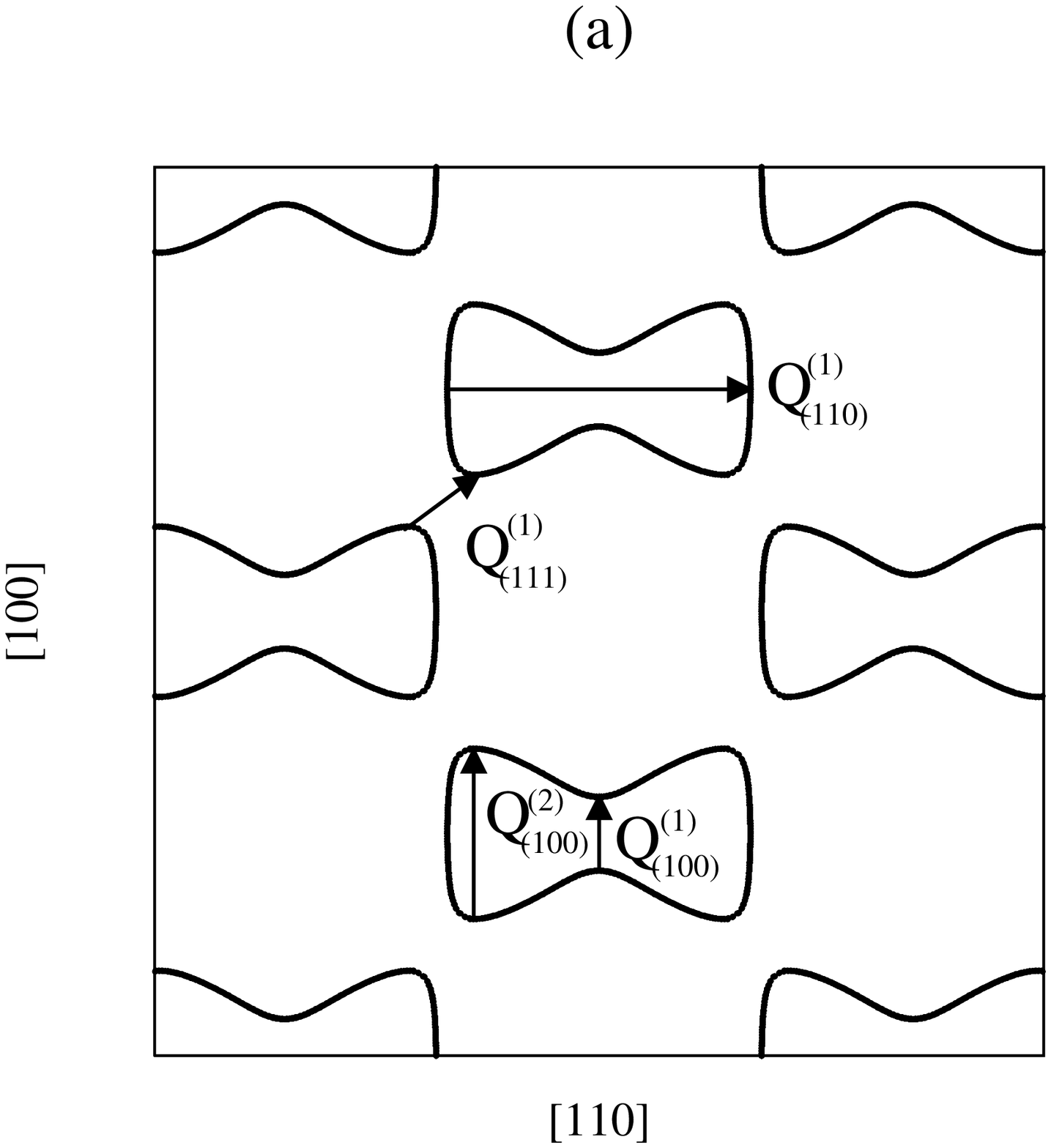,width=6.cm} &
\psfig{figure=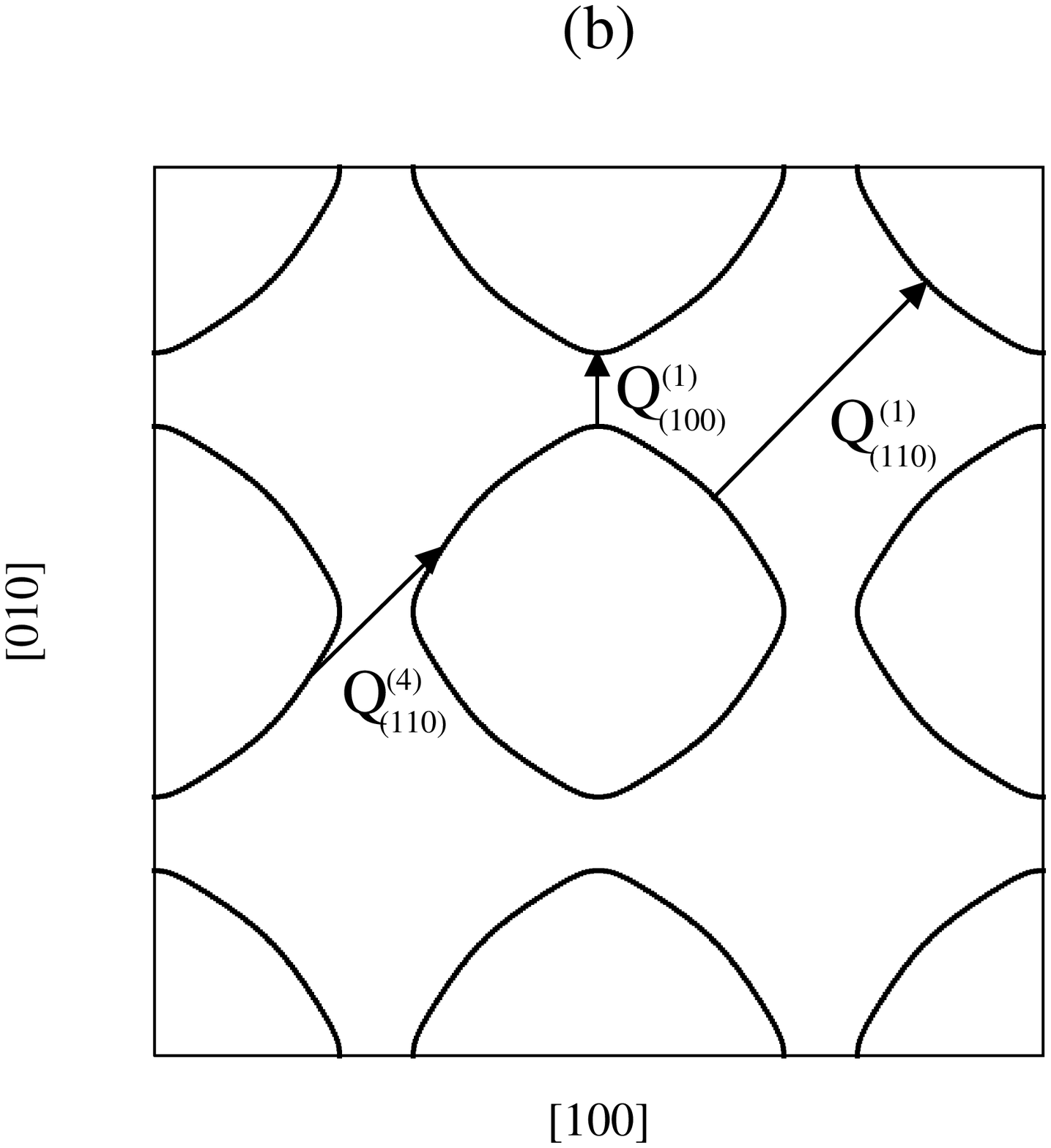,width=6.cm} \\
\multicolumn{2}{c}{\psfig{figure=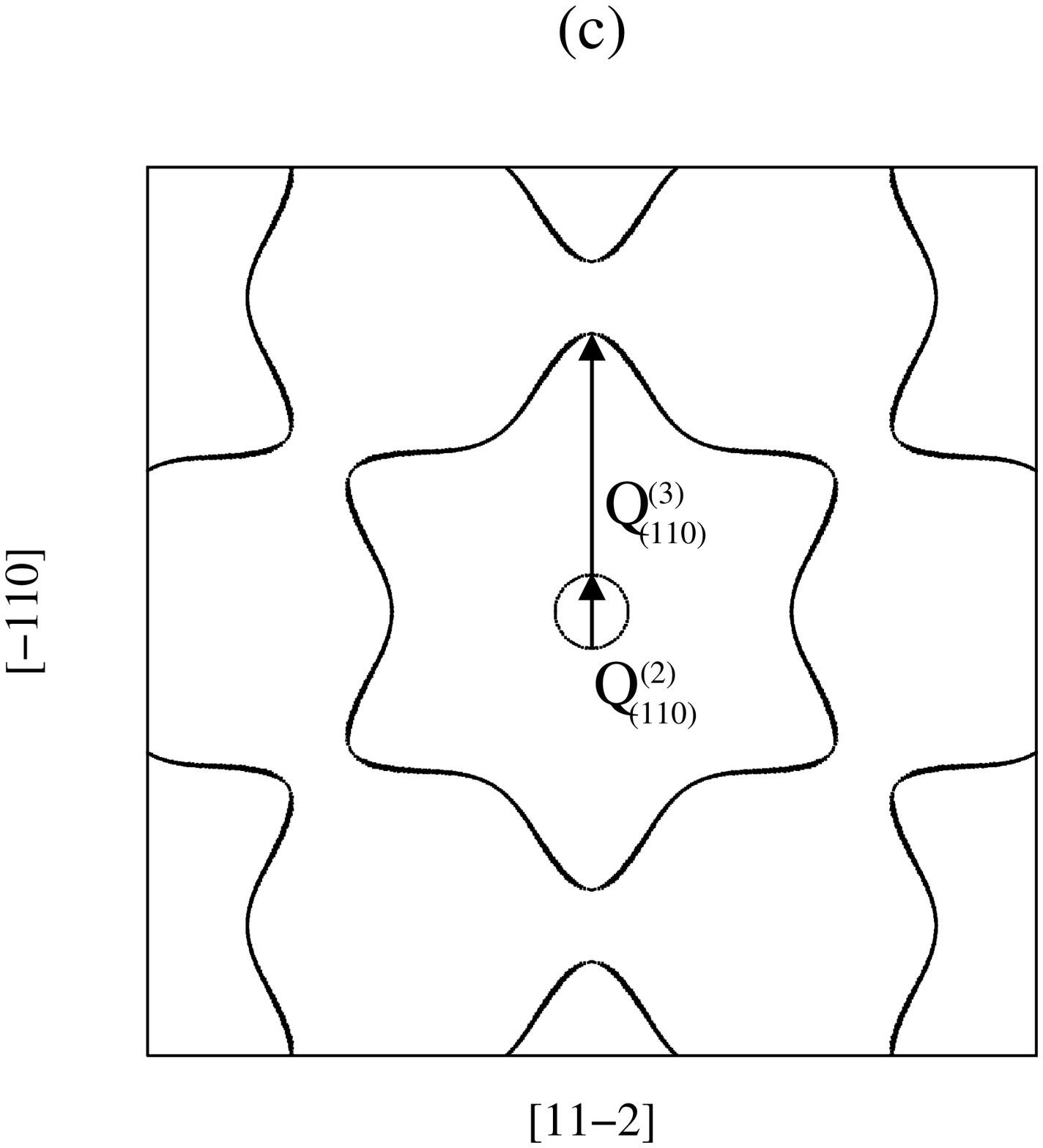,width=6.cm} } \\
\end{tabular}
}
\caption{Three different cuts of the Cu Fermi surface in the repeated zone
scheme with all the extremal vectors:
(a) perpendicular to  
the [1-10] direction at distance $\Delta k$=0 to the $\Gamma$ point,
(b) perpendicular to the [001] at distance $\Delta k$=0 and 
(c) perpendicular to the [111] at 
$\Delta k$=$\protect\sqrt{3} / 2$.\label{fig:cu_fs}}
\end{figure}

\section{ Evolution of the Fermi surface of Cu$_{(1-x)}$Ni$_x$  alloys with 
concentration.}

The FCC  Cu$_{(1-x)}$Ni$_x$ alloys are one of the best known examples for which
the rigid band model completely misconstrues the nature of the changes in the 
electronic
structure on alloying\cite{nato,faulkner}. Thus a direct experimental study of
the Fermi surface is of fundamental interest from the point of view of the
electronic structure of metallic alloys. 

\begin{figure}
\centerline{\begin{tabular}{cc}
\psfig{figure=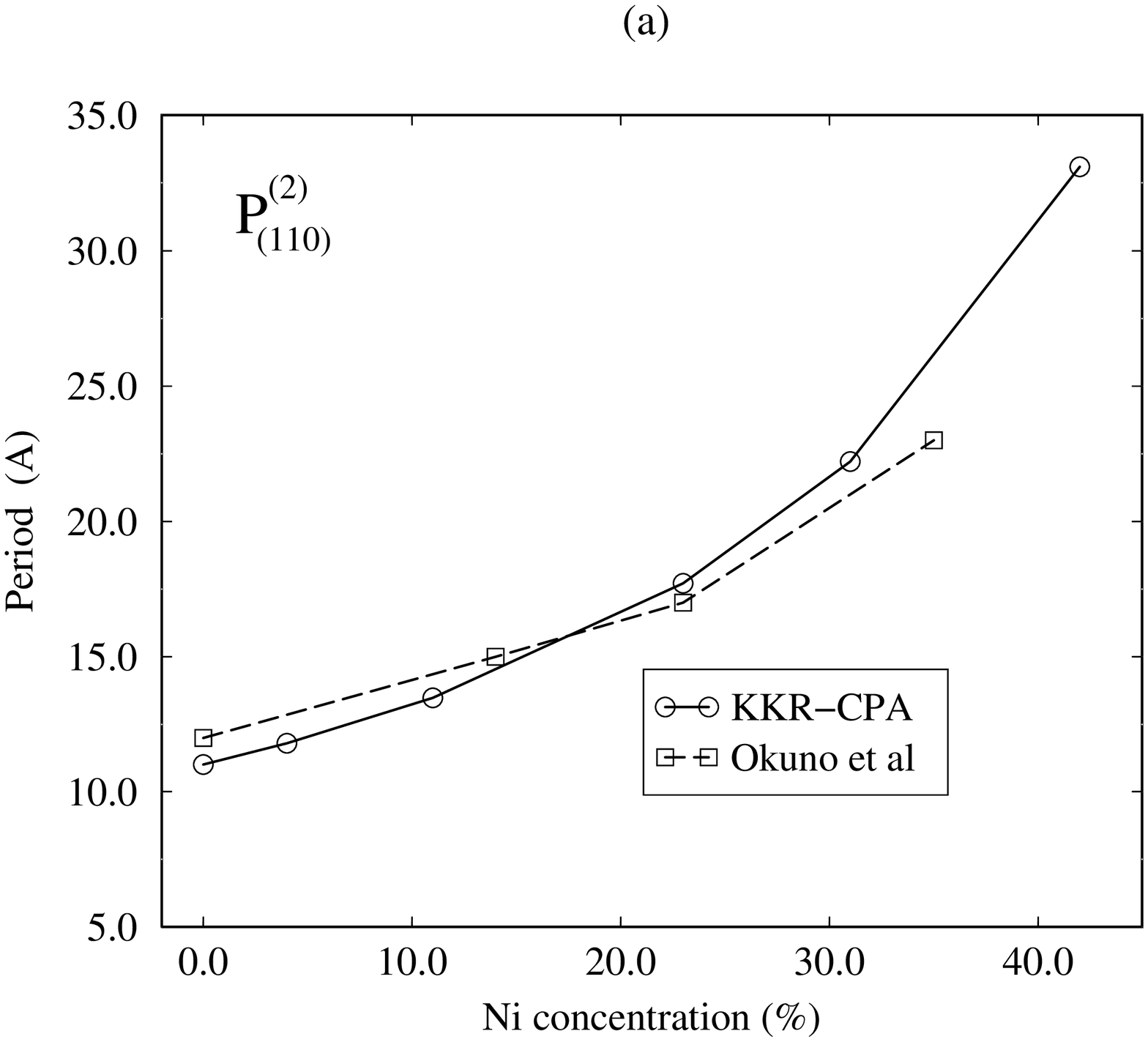,width=7 cm,angle=270} &
\psfig{figure=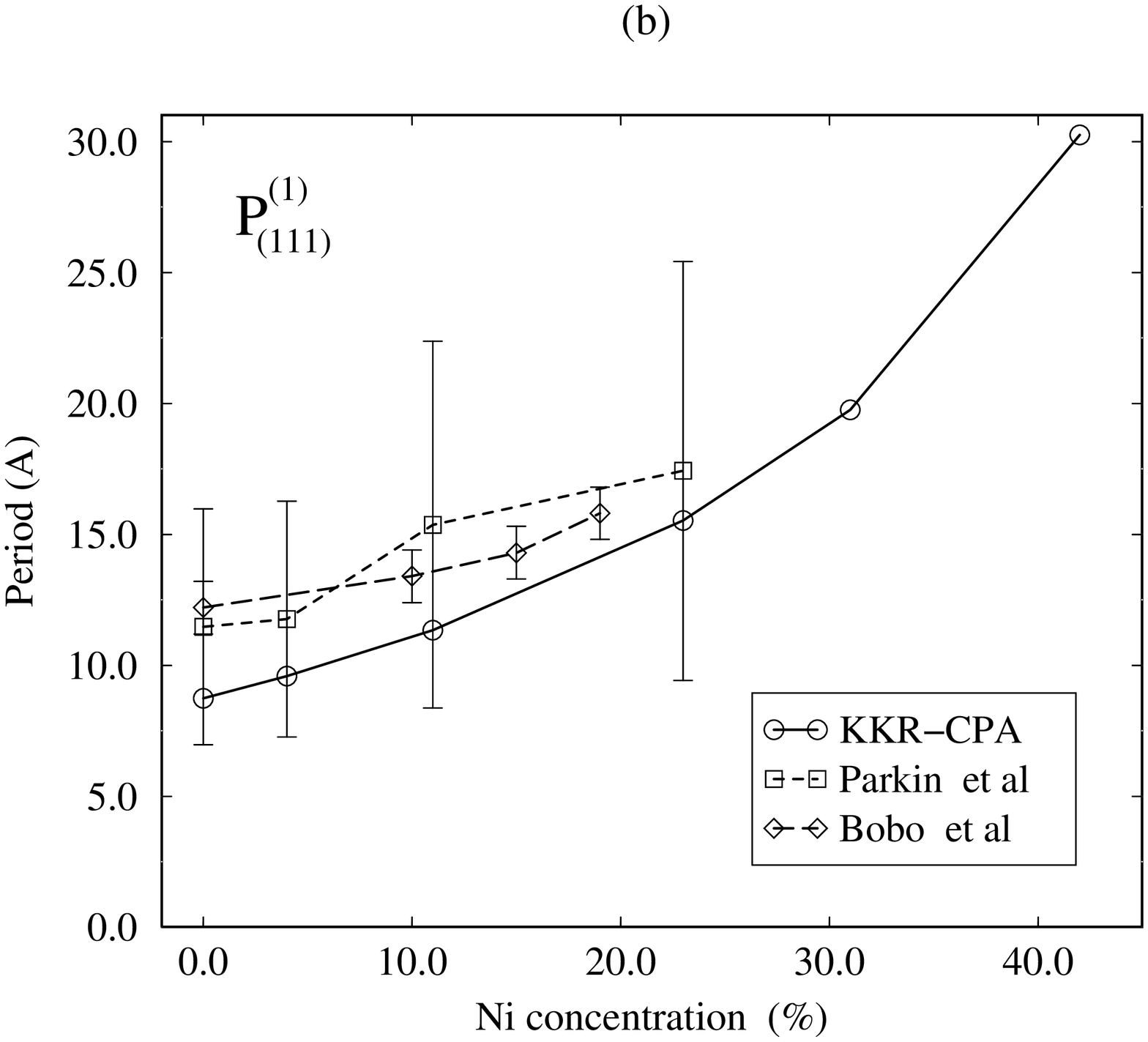,width=7 cm,angle=270} \\
\end{tabular}}
\caption{ Comparison of the calculated large periods as
functions of Ni concentration with the experiments of Bobo 
{\it et al}\protect\cite{bobo} and Parkin 
{\it et al}\protect\cite{parkin1} for the period 
$P_{(111)}^{(1)}=2\pi [Q_{(111)}^{(1)}]^{-1}$ (a) and
Okuno {\it et al}\protect\cite{okuno} for the period \label{fig:compa} 
$P_{(110)}^{(2)}=2\pi [Q_{(110)}^{(2)}]^{-1}$ (b).}
\end{figure}

Oscillations for this alloy systems have been observed in three separate
experiments. Parkin {\it et al}\cite{parkin1} and Bobo {\it et al}\cite{bobo}
studied a Co/Cu$_{(1-x)}$Ni$_x$/Co
system with (111) growth direction, while Okuno{\it et al}\cite{okuno} studied the same
system for the (110) orientation. The relatively long oscillation period 
($\sim$~10\AA)
observed in all of these experiments is believed to correspond to 
neck caliper vectors of the Cu-like Fermi surface of Cu$_{(1-x)}$Ni$_x$ alloys
for $x\leq 0.4$. In particular in the case of (110) direction the caliper vector
is the diameter of the neck itself, while in the case of (111) orientation
it spans the neck in an angle of 19.47$^o$ with respect to the neck plane. 
In fig.~\ref{fig:cu_fs} all the extremal
vectors for the (100), (110) and (111) directions are shown. The ones we already
mentioned for the (110) and (111) directions are the $Q_{(110)}^{(2)}$ and 
$Q_{(111)}^{(1)}$ respectively. In the refs.\cite{our1,our2} we have calculated 
the
concentration dependence of these extremal vectors using the KKR-CPA electronic
structure method\cite{nato,method} 
and have compared the predicted oscillation periods with the
experimental ones. We present that result also in fig.~\ref{fig:compa} for both
the (110) and (111) orientations. 

The decrease with concentration of the neck
diameter, resulting in an increase of the oscillation period, is in
qualitative agreement with the Rigid Band Model. Indeed, that decrease
comes from the fact that an electron-like neck of the Fermi surface such as the
one we are looking for is expected to shrink when electrons are removed from the
system (for instance by increasing the Ni concentration) but the calculated 
as well
as the observed shrinkage is more gentle than the predicted from the Rigid Band 
Model. The excellent agreement between our calculation and the experiment as
illustrated in fig.~\ref{fig:compa} is one more striking example of the 
success of the
CPA theory in binary alloy systems. The relatively simple Fermi surface of the 
Cu$_{(1-x)}$Ni$_x$ binary alloy system ($x<0.5$) makes it easy to examine 
whether the evolution of the Fermi surface with alloying is in agreement with
OMC measurements, and as is illustrated in fig.~\ref{fig:compa} that agreement
is excellent. In the next section a binary alloy system with much more
complicated Fermi surface is examined, namely the Cr$_{1-x}$V$_x$ binary alloy.

\section{ Which piece of the Fermi surface drives the long period oscillations
across Cr$_{(1-x)}$V$_x$ spacers?}

\begin{figure}
\centerline{\psfig{figure=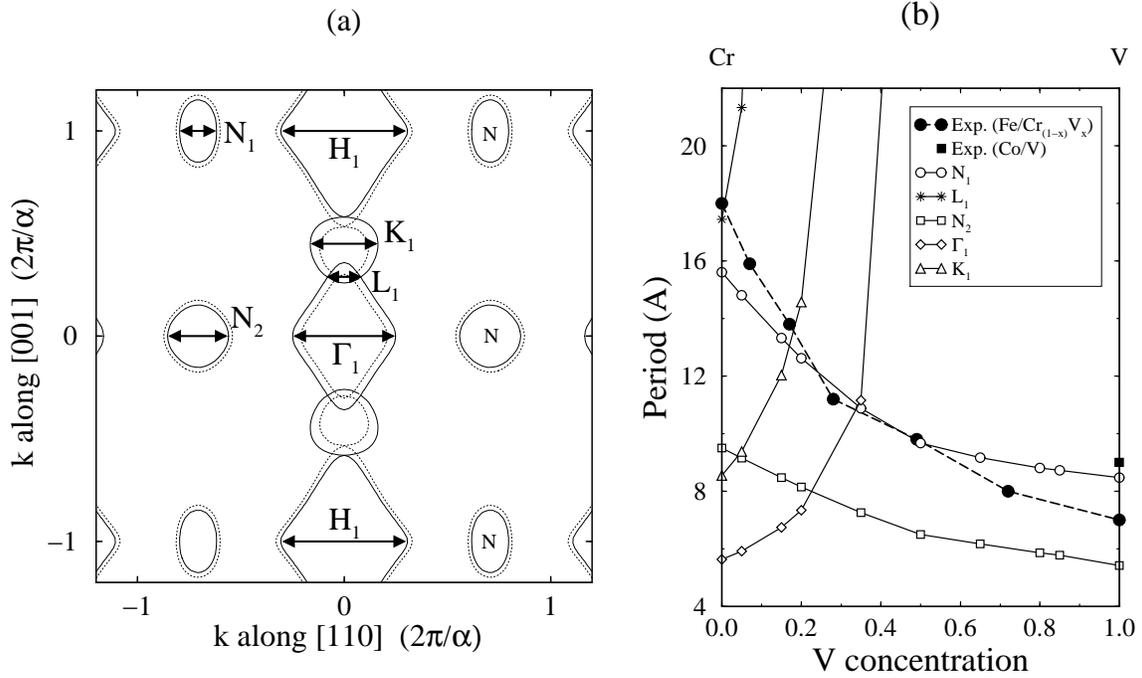,width=15cm}}
\caption{\label{fig:com_cr} (a) Cut of the pure Cr (solid)
and of Cr$_{0.85}$V$_{15}$ (dotted) Fermi surfaces, 
perpendicular to [1-1~0] direction
through $\Gamma$ point. 
(b) The Dependence of the oscillation periods (corresponding to the extremal
vectors shown in (a)) with V concentration. Experimental data from 
ref.~\protect\cite{schil2} for Fe/Cr$_{(1-x)}$V$_x$ and 
ref.~\protect\cite{parkin_co} for Co/V sandwiches are also included in (b) for 
comparison.  } 
\end{figure}

Although the long period oscillation for Fe/Cr/Fe ($\sim$18~\AA) was the first 
example of Oscillatory coupling discovered\cite{parkin0},
until recently the origin of that long period oscillation was the 
subject of an open
debate\cite{stiles,schilf_harr,koelling,dli,mirbt2,stiles2,tsetseris}. 
In the literature, 
the most popular candidate parts of the Fermi surface for being relevant
 are the electron-like lens of the 
Fermi surface and the N point centered hole-like pocket. A further point of
interest is that the long period oscillation appears to be unaffected by the
orientation of the specimen, at least for the (100), (110) and (211) directions,
leading to the conclusion that the coupling 
comes from a fairly isotropic region of the Fermi surface\cite{fullerton}. 
The extra complication that makes it extremely difficult to 
associate the oscillation period with an area of the  Fermi surface comes
from the fact that the Cr Fermi surface is fairly complicated with high degree
of nesting features arising from the d-states. 
Its rather difficult even to 
enumerate all the extremal vectors of the Fermi surface and of course the
procedure of just comparing the experimental periods with the modes of the 
Fourier transform of a total energy calculation totally fails in this case. Of
course the panacea would be to calculate the amplitudes of the individual 
oscillatory terms and reveal which terms are dominant. Such calculations have
been done using semi-empirical Tight Binding methods\cite{stiles,tsetseris}, and
although the size of the associated period is significantly smaller than the 
observed, they conclude that the origin of the oscillation is the N hole-like 
pocket for both the (100) and (110) orientations. Unfortunately,
other authors are drawing different conclusions\cite{koelling,fullerton}. 

In ref.~\cite{our_prl} we suggested that the evolution of the Fermi surface 
with alloying in the Cr spacer could give conclusive answer to the debate.
Indeed, there are experiments on the OMC across Cr$_{(1-x)}$V$_x$ as well as 
Cr$_{(1-x)}$Mn$_x$ spacers for poly-crystalline samples with (110) predominant
orientation\cite{schil2}. The idea is that if the origin of the oscillation is 
an electron-pocket then that pocket should shrink as V is added enhancing the 
size of the oscillation period. If on the other hand, the origin is a hole-like
pocket the period should decrease when V is added. The opposite apply in the
case of alloying with Mn. What the experiment shows is a monotonic 
decreases of the period with
V concentration and increase with Mn concentration, which is consistent with
the source of oscillation being a hole-like pocket. 
Of course the RBM is not
quantitatively correct in general, but it serves as a good qualitative picture.
Of course, our calculation is based on the KKR-CPA method 
and someone expects 
the agreement to be beyond the qualitative level.  
In fig.~\ref{fig:com_cr} the experimental period is shown  as a function of 
the V concentration along with 
the theoretical periods predicted from various extremal vectors of the 
alloy Fermi surface. As we see the period predicted from the N-hole-like
pocket is the only one which agrees quantitatively with experiment. 
Thus, our results strongly suggest that the source of the oscillations 
for both pure Cr as well as Cr$_{(1-x)}$V$_x$ spacers is the N-hole-like
pocket of the Fermi surface\cite{our_prl}.  For this 
particular case the rigid band model seem to agree quantitatively with the
more accurate CPA result, as has been shown by Koelling\cite{koelling1} who
used that model to draw similar conclusion to ours for the Cr spacer long 
period oscillation.

\begin{figure}
\centerline{\psfig{figure=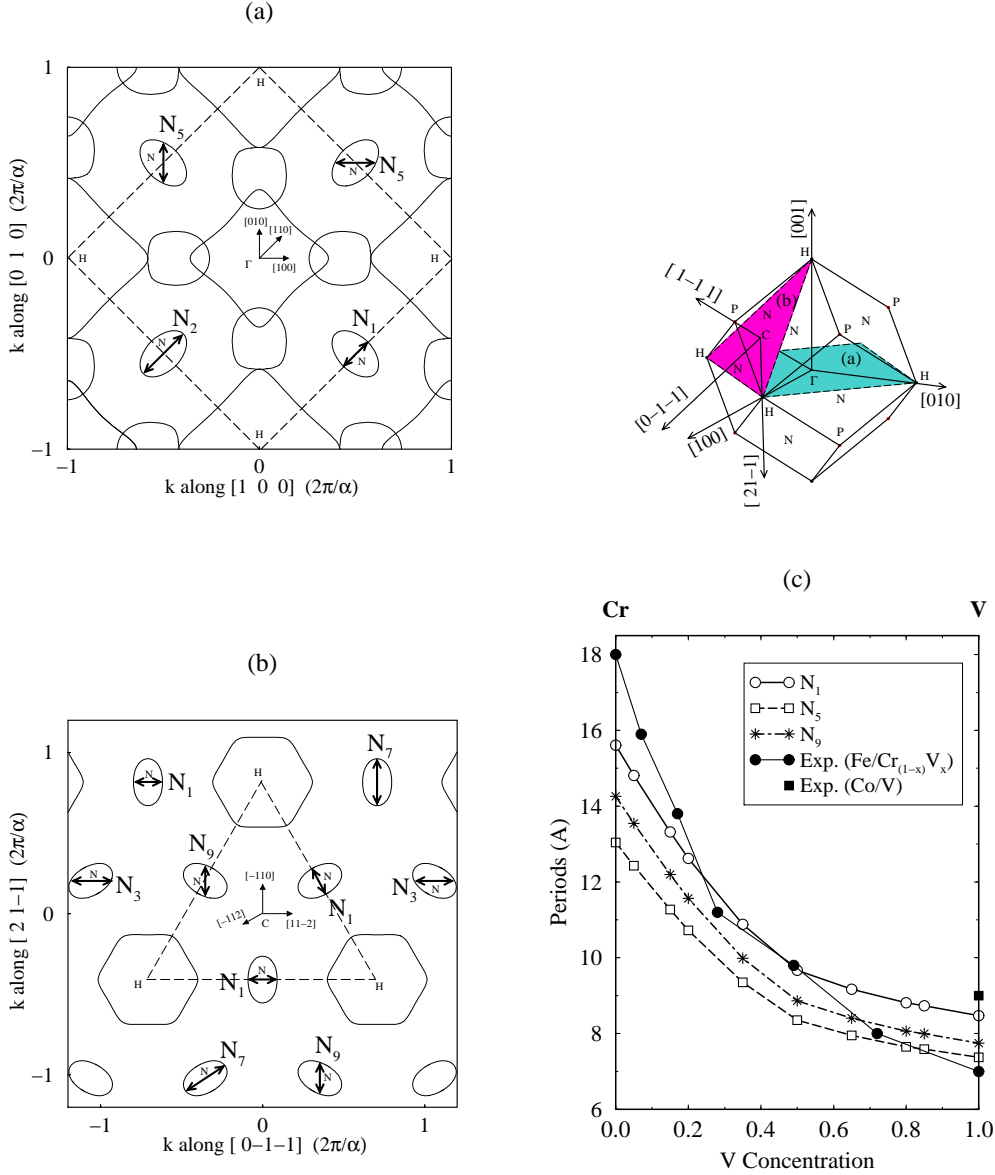,width=13cm}}
\caption{\label{fig:all_ellips}(a),(b) The Brillouin zone of BCC lattice with 
cuts of the Fermi volume by the two planes shown in the inset.
N$_1$, N$_2$, N$_3$ are the extremal vectors for the (110) direction,
while N$_4$, N$_5$ are those for the (100) and N$_6$, N$_7$, N$_8$, N$_9$
for the (211). The N$_4$ (not shown in (a) and (b) cuts) is the ellipsoid 
principal axis along the NP direction. N$_6$ and N$_8$ also are not shown 
in these cuts.
(c) The largest oscillation periods for each of the  (110), (100) and (211) 
directions corresponding to the  N$_1$, N$_5$ and N$_9$
extremal vectors respectively as functions of V concentration. 
The dependence on V concentration of the periods corresponding to the
rest of the extremal vectors is similar but the sizes of these periods are
significantly smaller than the ones plotted.
The experimental data of Parkin for Fe/Cr$_{(1-x)}$V$_x$~\protect\cite{schil2} 
and Co/V~\protect\cite{parkin_co} sandwiches 
are also included in (c) and refer to the (110) direction. 
}
\end{figure}

The lattice mismatch at the interfaces of the sandwich structures, is
one of the factors that could probably affect the agreement between 
theory and experiment in comparing the sizes of the oscillation periods to the
sizes  of the extremal vectors of the bulk Fermi surface\cite{our_vanc}. In the 
case of Cr as well as Cr$_{(1-x)}$V$_x$ spacers, that lattice mismatch is not 
important since the lattice constants of Cr, V, and Fe are very close to each
other, but in other cases like for example Fe/Mo/Fe or 
Fe/Cr$_{1-x}$Mo$_x$/Fe
alloy spacers the effect of lattice mismatch might be large enough to 
be ignored. Thus for instance in ref.~\cite{our_vanc}, we argue that it could be
the explanation for the discrepancy between the Rigid Band Model and the 
significant decrease with concentration been observed experimentally for
Fe/Cr$_{1-x}$Mo$_x$/Fe systems. The Rigid Band Model is been proved to be 
correct 
for the isoelectronic Cr, Mo and their alloys and is not consistent with the 
decrease in the oscillation period with concentration. We argue in
ref,~\cite{our_vanc} that the size of the effect of lattice mismatch is enough
to explain such a behavior, although someone needs to know the exact
geometry of the sandwich structure to draw a conclusive evidence. 

Having established the relation of the N-hole pocket and the OMC across 
Cr and Cr$_{(1-x)}$V$_x$ spacers the OMC is proven to be a powerful 
experimental technique for studying the geometry of that pocket and how it
evolves with concentration. In particular the N-hole pocket ellipsoid 
appears to grow isotropically with V concentration as is shown in 
fig.~\ref{fig:all_ellips}. In that figure the periods predicted form the 3 
smallest N pocket extremal vectors for the (100), (110) and (211) directions 
are shown as functions of V concentration. That 
does not appear to be the case in recent 2d-ACAR experiments\cite{alam} where a 
rotation of the N-hole pocket is observed with increasing concentration 
of Vanadium. The only experimental technique apart from 
2-d ACAR which could resolve this  extremely delicate feature of the 
Fermi surface appears to be the OMC. 

The long period oscillation across Cr$_{(1-x)}$V$_x$ spacers is an example 
in which the alloy theory
gives a conclusive answer, by continuity to an outstanding problem concerning 
the pure metal spacer system.

\section{\label{SEC:4}The exponential damping due to disorder scattering.}

\begin{figure}
\begin{tabular}{cc}
\psfig{figure=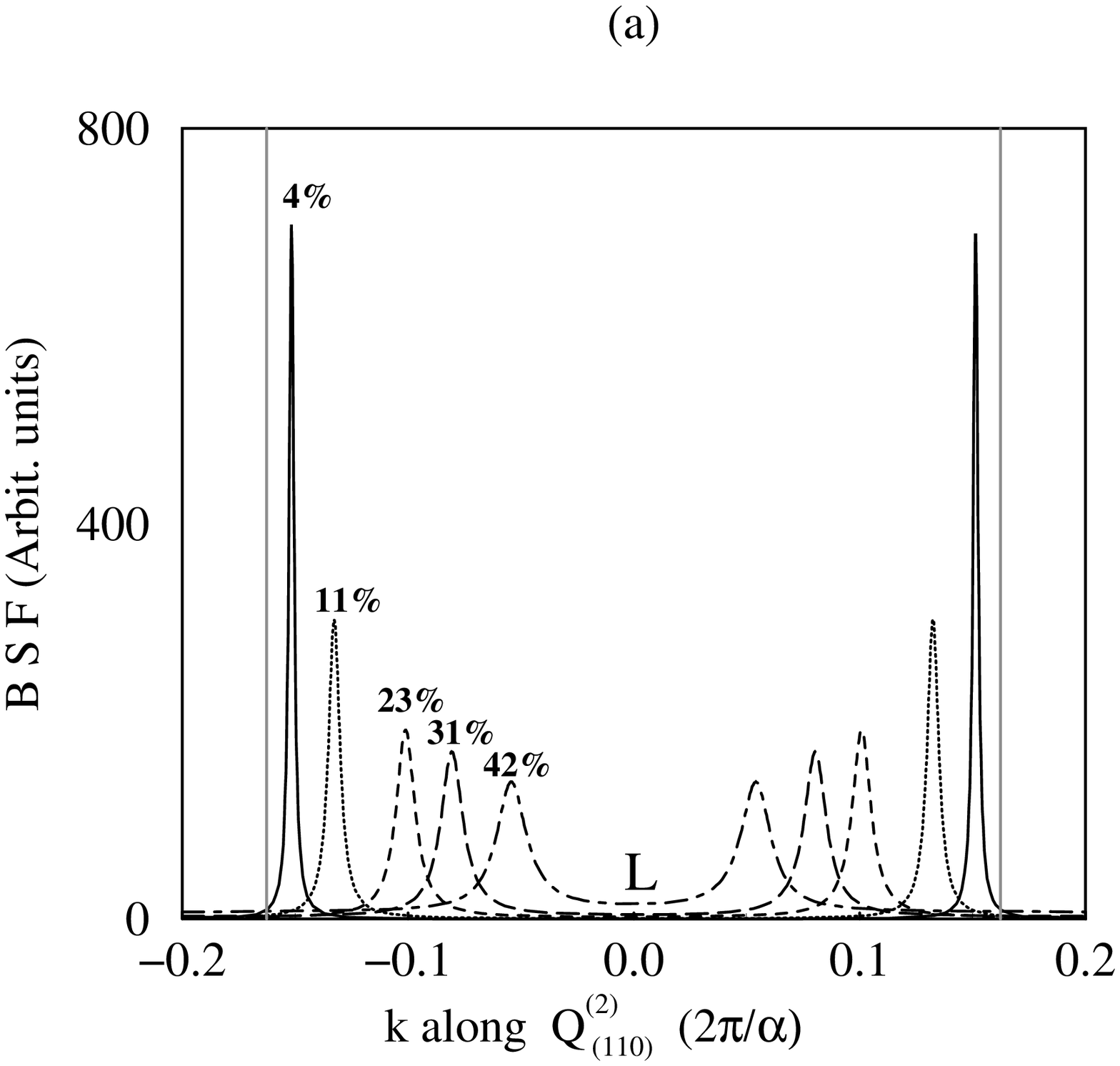,width=7cm,angle=0} &
\psfig{figure=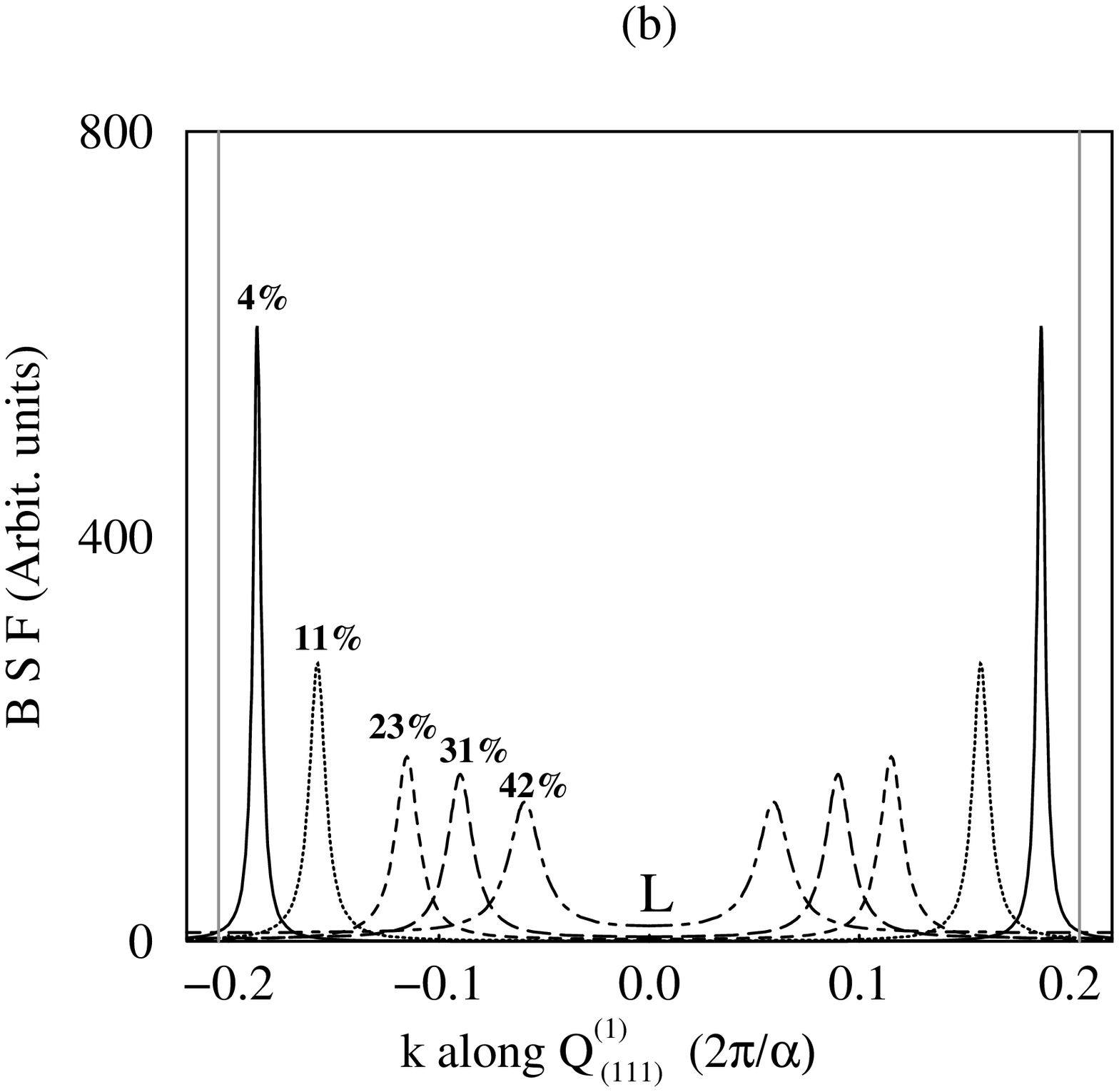,width=7cm,angle=0} \\
\psfig{figure=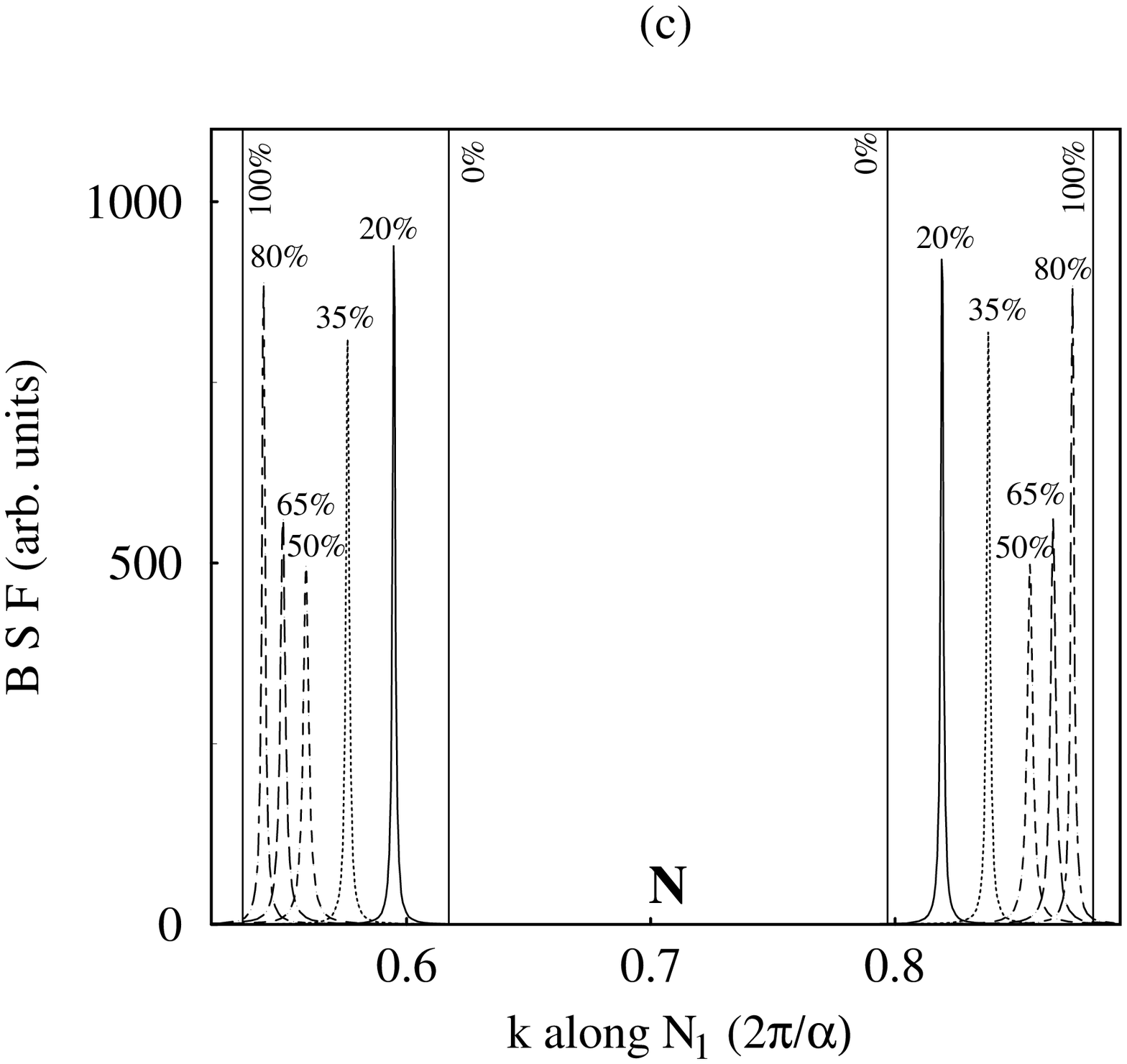,width=7 cm,angle=0} &
\psfig{figure=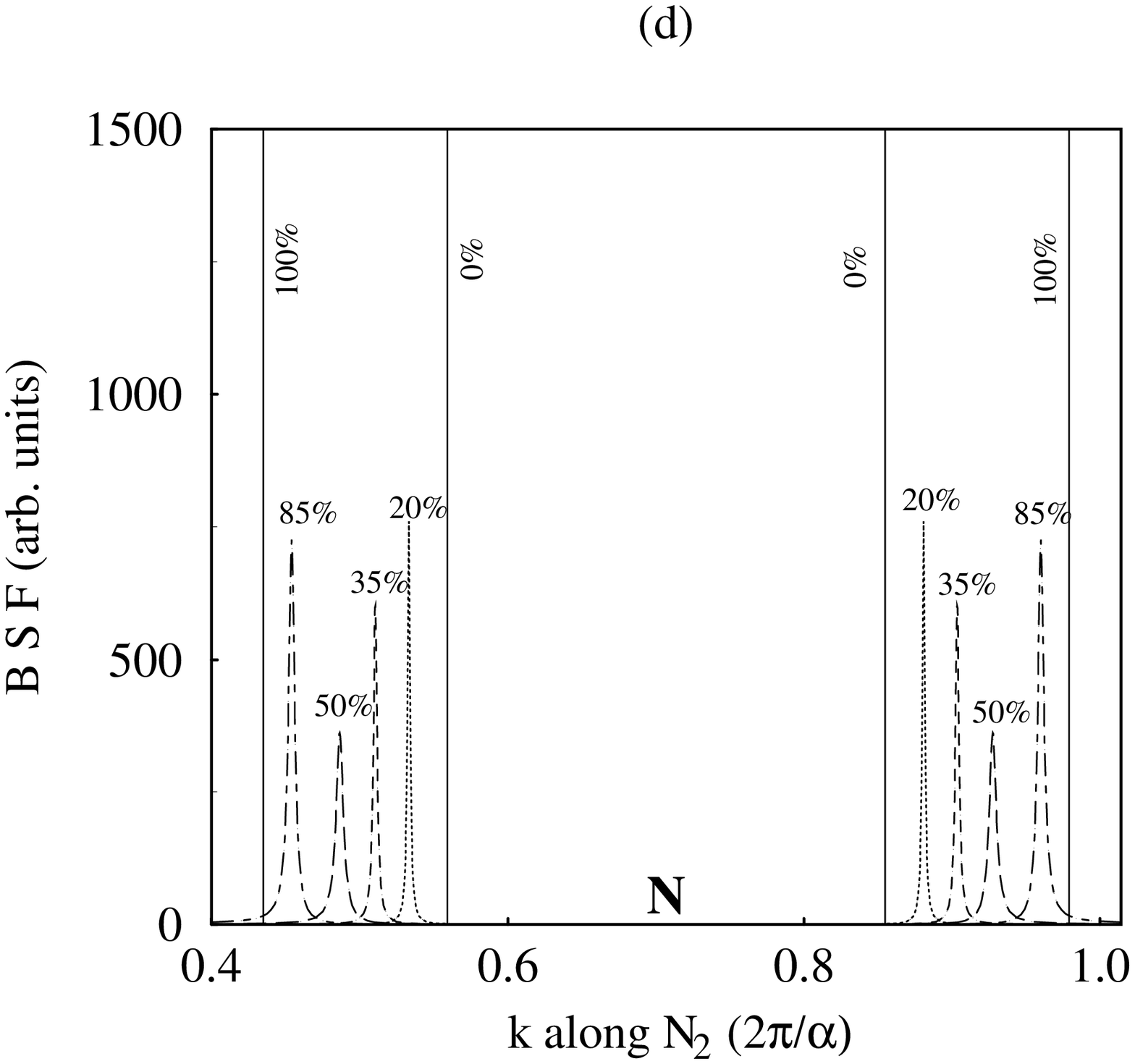,width=7 cm,angle=0} \\
\end{tabular}
\caption{\label{fig:bsf}The BSF along the direction of the extremal vectors 
$Q_{110}^{(2)}$ (a) and $Q_{111}^{(1)}$ (b) for Cu$_{(1-x)}$Ni$_x$ and
N$_1$ (c) and N$_2$ (d) for Cr$_{(1-x)}$V$_x$ for various concentrations.}
\end{figure}

As we have already mentioned an exponential damping of the OMC is present in
the case of disordered binary alloy spacers. The characteristic length of the
damping, {\it i.e.} the quantity $\Lambda_\nu$ in the eq.~(\ref{eq:2}) is
related to the 
coherence length of the quasi-particle states at the endpoints of the extremal
vector, which is a measure of the mean free path for these states. A convenient
quantity to describe the electronic structure of substitutionally disordered
systems such as the random binary alloys is the so called Bloch Spectral
Function (BSF) $A_B({\bf k},E)$ which is the number of states per Energy and 
wave length\cite{nato}. In the
case of pure metals that function is simply a sum of delta functions either as
function of $E$ at a constant wavevector $\bf k$, or as a function of the 
wavevector $\bf k$ for
constant value of the energy. For constant $E=E_f$ where $E_f$ is the Fermi
energy the positions of the peaks in k-space define the Fermi surface of the
metal. The k-space representation is still a good description of the electronic 
structure of the alloy although strictly speaking there is periodicity only on
the average. In
terms of the BSF, the fundamental difference is the peak lowering and broadening
being Lorentzian-like. Thus a Fermi surface for the alloy is still defined
through the position of the peaks but these peaks have a finite width, the
inverse of which defines the coherence lengths we mentioned above. In a simple 
theory for the OMC in the case of disordered spacer in ref~\cite{our1} we 
showed that 
\begin{equation}
\frac{1}{\Lambda_\nu}
= \Gamma^{(+)}_\nu - \Gamma^{(-)}_\nu 
\label{eq:mean_free_path}
\end{equation}
where $\Gamma^{(+)}$, $\Gamma^{(-)}$ are the widths of the BSF peaks along the
direction of the extremal vector with $(+)$ and $(-)$ labeling the two peaks at
the end-points of the extremal vector. Of course the Fermi surface is well
defined if the size of  $\Gamma^{(+)}_\nu$ 
$\Gamma^{(-)}_\nu$ is small compared to $Q_\nu$, {\it i.e.} the size of the
extremal vector itself. In that case of course, $\Lambda_\nu$ is large compared
to the oscillation period $P_\nu$, {\it i.e.} no damping is observed within
the first few oscillation periods. 

The obvious question in the light of the above discussion is of course how broad
are the spectral functions for the cases of extremal vectors we considered 
above for the Cu$_{(1-x)}$Ni$_x$ and Cr$_{(1-x)}$V$_x$ Fermi 
surfaces? In fig~\ref{fig:bsf} we show the BSF along these extremal vectors.
We see that for Cu$_{(1-x)}$Ni$_x$ at concentrations of the
order $x\approx 0.5$, where an electronic topological transition (ETT) takes 
place\cite{beniamino}, in 
order the widths 
become comparable with the size of the extremal vectors. On the other hand for
Cr$_{(1-x)}$V$_x$ the extremal wave vector size is always very large compared 
to the width of the peaks. The BSF peaks for the N-hole pocket appear to be the
sharpest for the whole Fermi surface. Although ETT occurs 
in other parts of the Fermi surface, the N-hole pockets are robust in alloying,
with only its total volume increased as more V is added. Thus, in both 
Cu$_{(1-x)}$Ni$_x$ ($x\leq 0.4$) and Cr$_{(1-x)}$V$_x$ for the whole range of V
concentration, no significant damping is expected agreement with
the experiments on these two systems\cite{parkin1,bobo,okuno,schil2}.

It would be interesting if an alloy spacer system was found with a strong
topological
transition happening in the area of the extremal vector at some value of the 
concentration $x_o$. The exponential damping of the OMC for values of $x$ close 
to $x_o$ could be measured and the dramatic prediction of the theory is that
no oscillation would be 
observed for the concentration where the ETT occurs. 
Such an experiment would be 
a direct observation of the ETT. Moreover, the characteristic length of 
the damping for $x$ close to but not equal to $x_o$ would be a direct 
measurement of the coherence length at particular points of the Fermi surface,
{\it i.e.} the endpoints of the extremal vector. Of course such measurements
require further refinement of the experimental techniques used for measuring 
the OMC. 
 
\section{\label{SEC:5}Calculations of the amplitudes $A_\nu$ and phases 
$\phi_\nu$.}

\begin{figure}
\centerline{\begin{tabular}{lr}
\psfig{figure=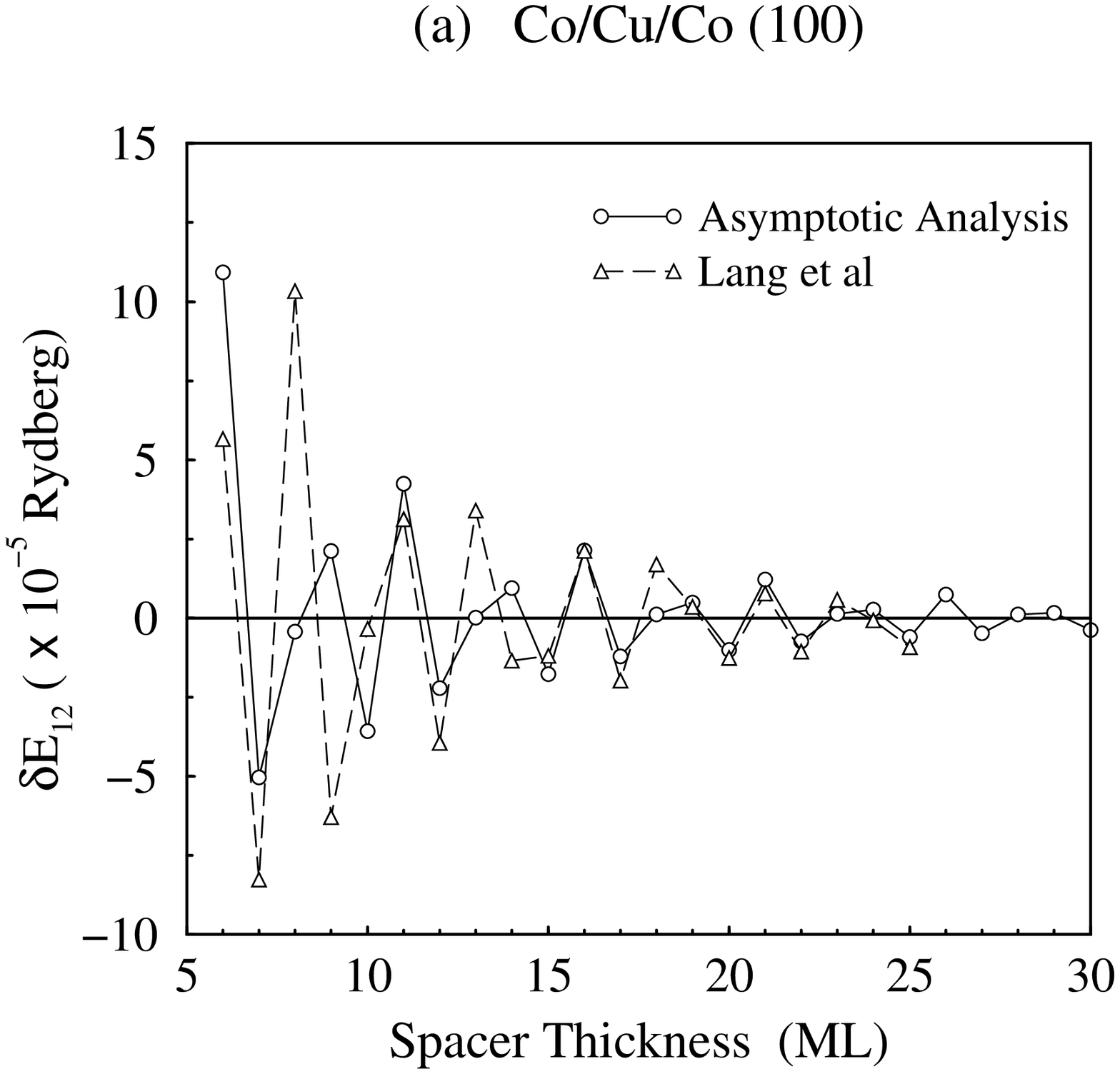,width=7.5cm} \ \ & 
\ \ \psfig{figure=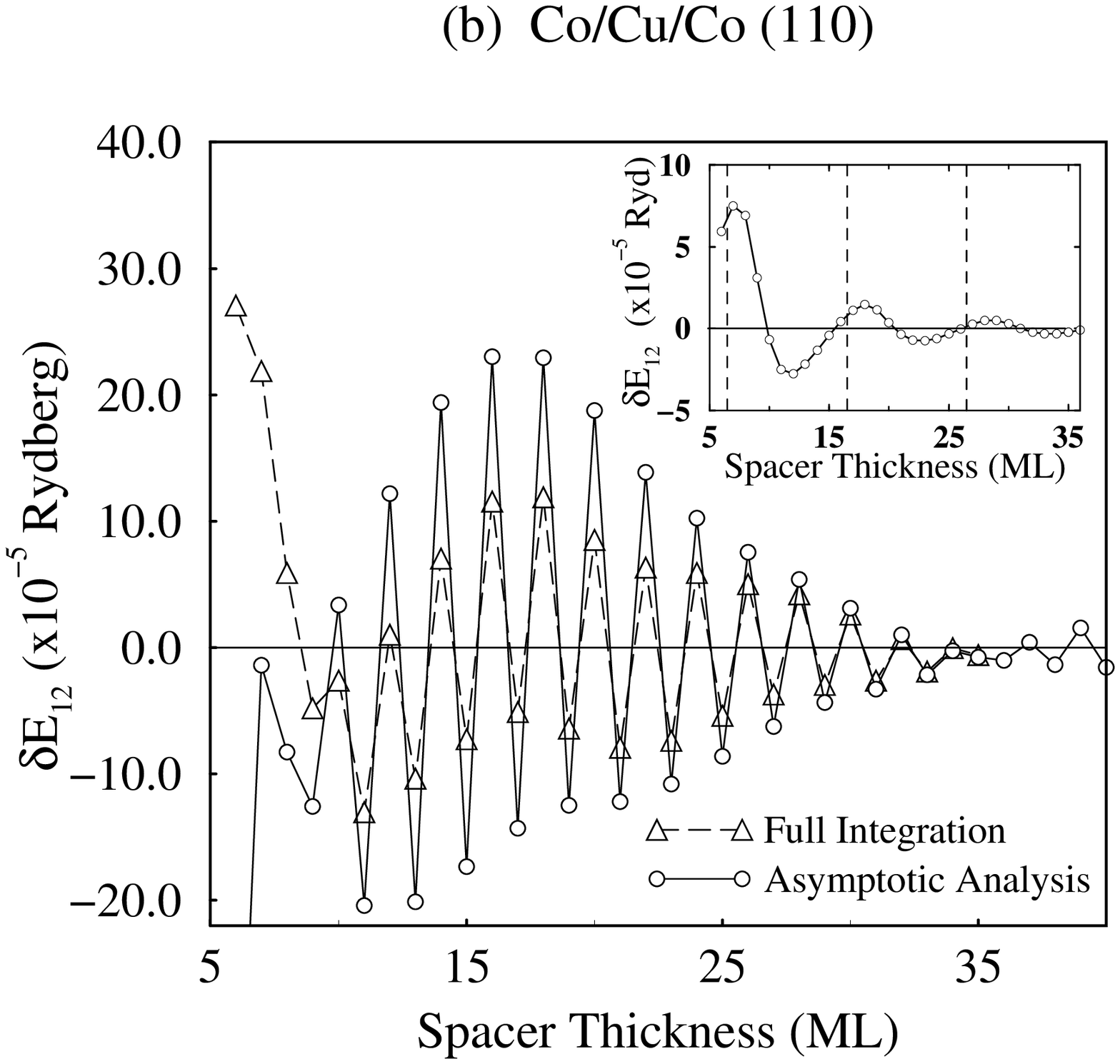,width=7.5cm} \\
\multicolumn{2}{c}{\psfig{figure=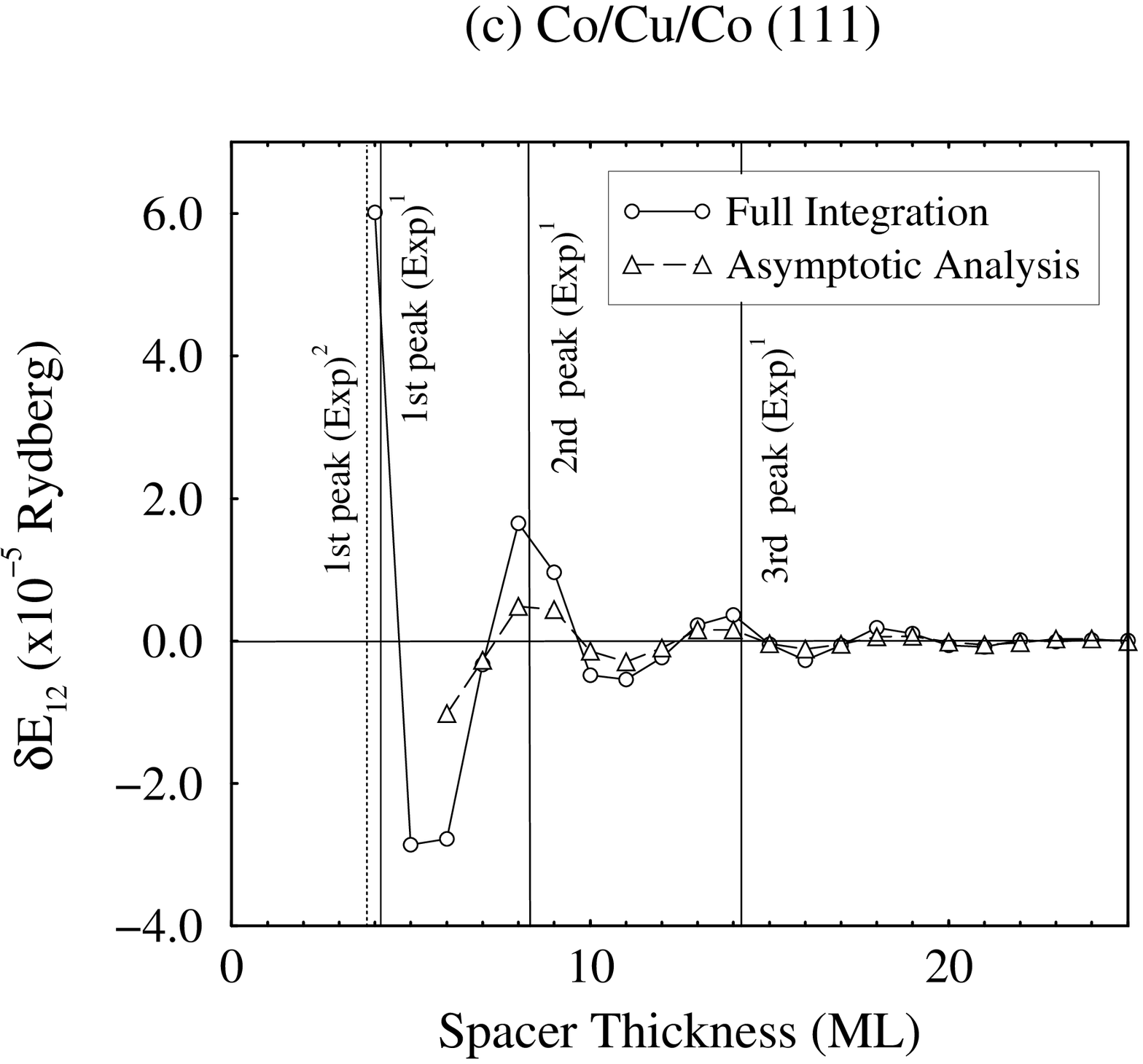,width=7.5cm}} \\
\end{tabular}
}
\caption{\label{fig:ampli} The calculated OMC for the Co/Cu/Co structure as
function of the spacer thickness:
(a) The asymptotic analysis result for the (100) orientation (solid line), 
compared with the total energy calculation result of Lang 
{\it et al}\protect\cite{lang}. (b) The OMC for (110) for both asymptotic
analysis (sum of two contributions, originating from 
$Q_{(110)}^{(1)}$ and $Q_{(110)}^{(2)}$ in fig.~\protect\ref{fig:cu_fs}) 
and full integration calculation for the (110) orientation.
In the inset the contribution of the Large period oscillation (originating from
$Q_{(110)}^{(2)}$) is plotted
with the vertical lines indicating the positions of the AF peaks found in 
experiment\protect\cite{johnson2}. (c) The OMC for both full integration and 
asymptotic analysis 
calculations for the (111) orientation. 
The vertical lines again indicate the 
positions of the AF peaks as found in the experiments of 
Johnson {\it et al}\protect\cite{johnson2} (Exp)$^1$ and Parkin 
{\it et al}\protect\cite{parkinn} (Exp)$^2$.}
\end{figure}

The asymptotic analysis leading to a theory of the amplitudes $A_\nu$ and the
phases $\phi_\nu$ has been carried out within a number of approaches to the 
problem of calculating the distortions in the electronic structure of the spacer
by the magnetic layers\cite{our_phil}. In particular, it was 
carried out for the powerfull,
fully first principles screened KKR method\cite{szugnyogh}. 
Quantitative calculation of
the asymptotic formulas are compared with the results of full total energy
calculations and experiments in fig.~\ref{fig:ampli}. 
Similar calculations for alloy spacers are
in progress. 

As can be seen in fig.~\ref{fig:ampli} the agreement between our calculated 
amplitudes and phases with both theoretical \cite{lang} and experimental 
results\cite{johnson2,parkinn}
for the OMC across Co/Cu/Co is remarkable for all the orientations. In
particular there is very good quantitative agreement between our 
results and Lang {\it et al}\cite{lang} in the amplitudes as seen in 
fig.~\ref{fig:ampli}a
for the (100) orientation. Furthermore, there is excellent agreement with the
experiments of refs.~\cite{johnson2,parkinn} concerning the phases and periods,
{\it i.e.} the positions of AF peaks, as seen in the  inset of 
fig.~\ref{fig:ampli}b and in fig.~\ref{fig:ampli}c. 
There is only agreement in the order of magnitude with
experiments concerning the size of the interaction. For example, for the 
(100) orientation Johnson {\it et al}\cite{johnson2} have measured 0.4~mJ/m$^2$ 
for an average Cu thickness of 6.7 ML, while we find 3.3~mJ/m$^2$ and 
-1.0~mJ/m$^2$ for 6 and 7 ML respectively. For the (111) 
Johnson {\it et al}\cite{johnson2} measured 1.1~mJ/m$^2$ for 4 ML of Cu
thickness, while our calculated value for the same thickness is 2.1~mJ/m$^2$.
The agreement with experiments concerning the amplitudes, although it is
restricted to the order of magnitude so far,
is at least a convincing evidence that our models have picked up the  
effect which is also seen by experiments.

In conclusion, we note that the asymptotic formulas such as the one 
we derived in \cite{our_phil} may play a
role in interpreting OMC experiments similar to that of the Lifshitz-Kosevich
semi-classical formula, used to deducing the Fermi surface geometry from the
measurement of the de Haas van Alphen oscillations.

\end{document}